\documentclass[aps,prd,reprint,twocolumn,superscriptaddress,showpacs]{revtex4-1}
\usepackage{graphicx}
\usepackage{mathrsfs}
\usepackage{bm}
\usepackage{amsmath}
\usepackage{dcolumn}
\usepackage{epstopdf}
\usepackage{dsfont}
\usepackage{amssymb}
\usepackage{tabularx}
\usepackage{array}
\usepackage{float}
\usepackage{color}
\usepackage{epstopdf}
\usepackage{mathrsfs}
\usepackage[colorlinks, linkcolor=blue,anchorcolor=blue,citecolor=blue,urlcolor=blue]{hyperref}

\begin{document}
\title{Two-dimensional antiferromagnetic Dirac fermions in monolayer TaCoTe$_2$ }

\author{Si Li}
\affiliation{School of Physics and Electronics, Hunan Normal University, Changsha, Hunan 410081, China}
\affiliation{Key Lab of Advanced Optoelectronic Quantum Architecture and Measurement (MOE),
Beijing Key Lab of Nanophotonics $\&$ Ultrafine Optoelectronic Systems, and School of Physics,
Beijing Institute of Technology, Beijing 100081, China}
\affiliation{Research Laboratory for Quantum Materials, Singapore University of Technology and Design, Singapore 487372, Singapore}

\author{Ying Liu}
\affiliation{Research Laboratory for Quantum Materials, Singapore University of Technology and Design, Singapore 487372, Singapore}
\affiliation{Institute for Theoretical Physics and Astrophysics, University of W\"{u}rzburg, W\"{u}rzburg 97074, Germany}

\author{Zhi-Ming Yu}
\affiliation{Research Laboratory for Quantum Materials, Singapore University of Technology and Design, Singapore 487372, Singapore}

\author{Yalong Jiao}
\affiliation{Research Laboratory for Quantum Materials, Singapore University of Technology and Design, Singapore 487372, Singapore}

\author{Shan Guan}
\affiliation{State Key Laboratory of Superlattices and Microstructures, Institute of Semiconductors, Chinese Academy of Sciences, Beijing 100083, China}

\author {Xian-Lei Sheng}
\affiliation{Research Laboratory for Quantum Materials, Singapore University of Technology and Design, Singapore 487372, Singapore}
\affiliation{Department of Physics, Key Laboratory of Micro-nano Measurement-Manipulation and Physics (MOE), Beihang University, Beijing 100191, China}

\author{Yugui Yao}\email{ygyao@bit.edu.cn}
\affiliation{Key Lab of Advanced Optoelectronic Quantum Architecture and Measurement (MOE),
Beijing Key Lab of Nanophotonics $\&$ Ultrafine Optoelectronic Systems, and School of Physics,
Beijing Institute of Technology, Beijing 100081, China}

\author{Shengyuan A. Yang}\email{shengyuan\_yang@sutd.edu.sg}
\affiliation{Research Laboratory for Quantum Materials, Singapore University of Technology and Design, Singapore 487372, Singapore}
\affiliation{Center for Quantum Transport and Thermal Energy Science, School of Physics and Technology, Nanjing Normal University, Nanjing 210023, China}

\begin{abstract}
Dirac point in two-dimensional (2D) materials has been a fascinating subject of research. Recently, it has been theoretically predicted that Dirac point may also be stabilized in 2D magnetic systems. However, it remains a challenge to identify concrete 2D materials which host such magnetic Dirac point. Here, based on first-principles calculations and theoretical analysis, we propose a stable 2D material, the monolayers TaCoTe$_2$, as an antiferromagnetic (AFM) 2D Dirac material. We show that it has an AFM ground state with an out-of-plane N\'{e}el vector.
It hosts a pair of 2D AFM Dirac points on the Fermi level in the absence of spin-orbit coupling (SOC). When the SOC is considered, a small gap is opened at the original Dirac points. Meanwhile, another pair of Dirac points appear on the Brillouin zone boundary below the Fermi level, which are robust under SOC and have a type-II dispersion. Such a type-II AFM Dirac point has not been observed before. We further show that the location of this Dirac point as well as its dispersion type can be controlled by tuning the N\'{e}el vector orientation.

\end{abstract}

\maketitle
\section{Introduction}

Since the discovery of graphene~\cite{novoselov2004}, two-dimensional (2D) materials have been attracting tremendous interest in the past decade. Many of the peculiarity properties of graphene can be ascribed to its peculiar Dirac-cone-type band structure~\cite{neto2009}, in which two bands cross at so-called Dirac points with linear dispersion at the Fermi level. Around the Dirac points, the low-energy electrons behave like relativistic massless Dirac fermions in 2D and exhibit properties distinct from the usual Schr\"{o}dinger fermions.

Inspired by graphene, there has been a continuing effort to search for 2D Dirac fermions in new materials and to predict new variants.
A number of 2D materials, such as silicene~\cite{cahangirov2009,liu2011}, germanene~\cite{cahangirov2009,liu2015multiple}, graphyne~\cite{malko2012}, 2D carbon and boron allotropes~\cite{xu2014,zhou2014,ma2016,jiao2016}, and group-Va monolayers~\cite{lu2016,kim2015}, have been predicted to host 2D Dirac points, but these points (including those in graphene) are unstable under spin-orbit coupling (SOC). 2D Dirac points robust against SOC, known as 2D spin-orbit Dirac points, have been proposed by Young and Kane~\cite{Young2015a}, and have been revealed in realistic 2D materials such as monolayer HfGeTe family~\cite{Guan2017} and monolayer $X_3$SiTe$_6$ ($X$= Ta, Nb)~\cite{li2018nonsym}. Recently, such spin-orbit Dirac points have been directly mapped out in the angle-resolved photoemission spectroscopy (ARPES) experiment on monolayer bismuthene with a black phosphorene type structure~\cite{kowalczyk2019}.

Generally, the stability of 2D Dirac points requires symmetry protection. The magnetic ordering breaks the time reversal ($\mathcal{T}$) symmetry and certain crystalline symmetries, and hence may destroy the Dirac points.
Therefore, most discovered 2D Dirac materials are nonmagnetic.
Recently, it has been discovered that Dirac fermions can be achieved in an antiferromagnetic (AFM) system, and the concept of
2D AFM Dirac semimetal has been proposed in theory~\cite{Wang2017,Young2017}. A few real materials, including monolayer FeSe~\cite{Young2017}, $X$Fe$_2$As$_2$ ($X$= Ba, Sr)~\cite{chen2017two}, and monolayer Zr$_2$Si~\cite{Shao2018} have been reported to be the possible candidates. However, the number of candidates is still very limited, and the proposed materials also have their own shortcomings. For example, the Dirac points in monolayer Zr$_2$Si are unstable under SOC~\cite{Shao2018}. For the monolayer FeSe, its Dirac point is away from the Fermi level~\cite{Young2017}. As for $X$Fe$_2$As$_2$ ($X$= Ba, Sr), it actually realizes quasi-2D Dirac fermions in a three-dimensional (3D) material~\cite{chen2017two}, hence lacks the great tunability of 2D materials. Thus, it is much desired to explore more realistic materials that can realize 2D magnetic Dirac fermions.

In this work, based on first-principle calculations and theoretical analysis, we reveal the monolayer TaCoTe$_2$ as a 2D magnetic Dirac material.
Bulk TaCoTe$_2$ is an existing layered magnetic material. We demonstrate that TaCoTe$_2$  remains stable in the monolayer form, and the calculated exfoliation energy is low (less than MoS$_2$), suggesting that the monolayer can be easily exfoliated from the bulk. We find that monolayer  TaCoTe$_2$ has an AFM ground state, with a high N\'{e}el temperature about 310 K.
In the absence of SOC, it hosts a pair of Dirac points at the Fermi level in the AFM state, protected by a glide mirror symmetry.
When SOC is included, a small gap about 60 meV will be opened at the original Dirac points, and the low-energy electrons become 2D massive Dirac fermions. Meanwhile, interestingly, there emerge another pair of magnetic Dirac points below the Fermi level, which are robust against SOC. Moreover, the two points exhibit a type-II dispersion, \emph{i.e.}, their Dirac cones are completely tipped over. To our knowledge, such type-II magnetic Dirac point has not been observed before. In addition, we show that the magnetic Dirac points in TaCoTe$_2$ have an interesting interplay with the orientation of the N\'{e}el vector. Our finding
provides a concrete material platform for the fundamental research of 2D magnetic Dirac fermions as well as for promising spintronics applications.

\section{CRYSTAL STRUCTURE}

\begin{figure}[t]
\includegraphics[width=8.8cm]{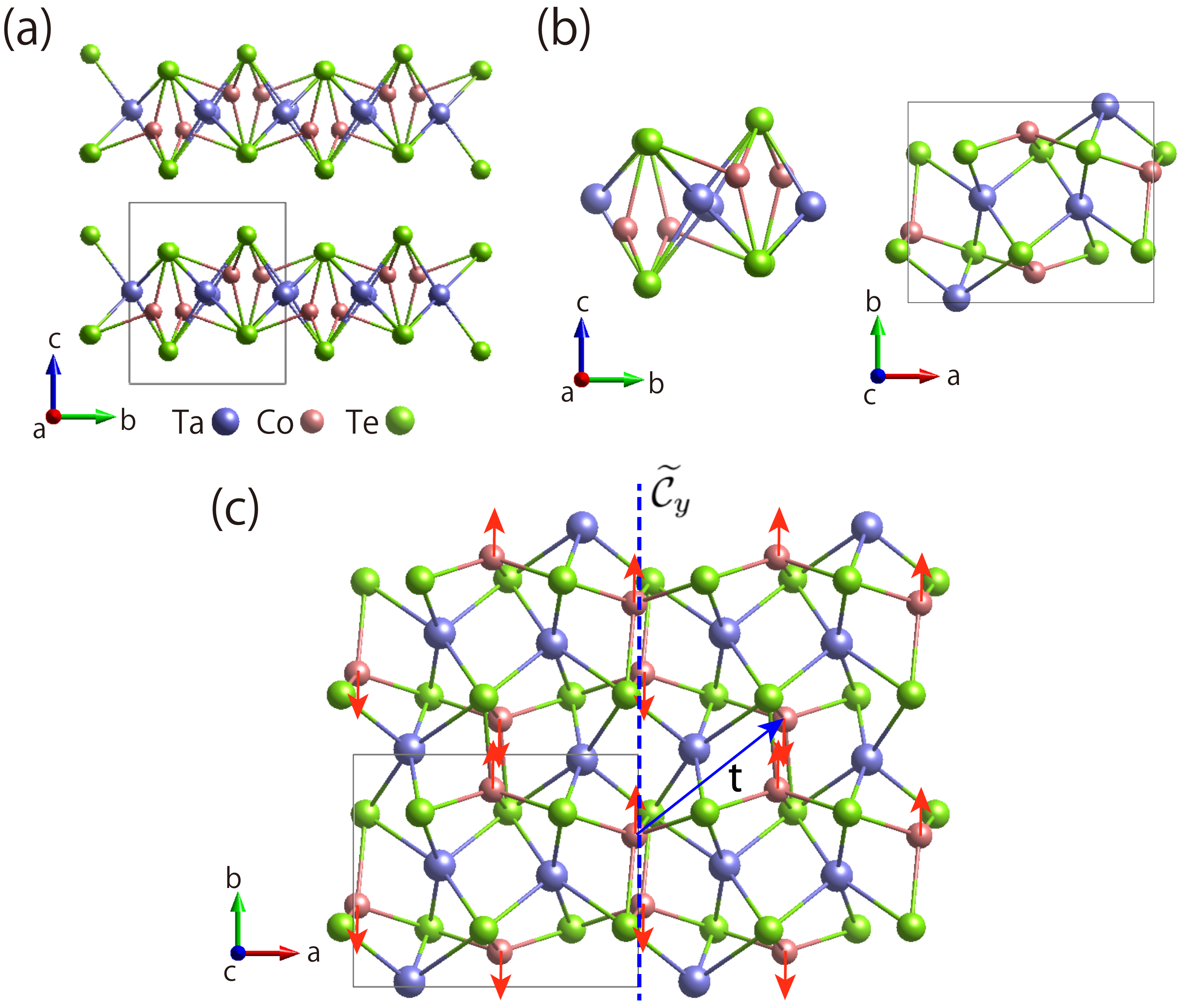}
\caption{(a) Crystal structure of the bulk TaCoTe$_2$. The primitive cell is shown with the solid line. (b) Side view and top view of the monolayer TaCoTe$_2$. (c) Illustration of the screw rotation symmetry $\widetilde{\mathcal{C}}_{y}: (x,y,z)\rightarrow (-x+\frac{1}{2},y+\frac{1}{2},-z)$. The blue	dashed line shows for the rotation axis, and ${\bf t}=(\frac{1}{2},\frac{1}{2},0)$ is the half translation along the [110] direction. Note that the red arrows represent the magnetic moments along the $z$ direction ($c$ axis).}
\label{fig1}
\end{figure}

The bulk TaCoTe$_2$ belongs to the ternary tellurides with a layered structure~\cite{li1993x} [see Fig.~\ref{fig1}(a)]. It has been synthesized from the component elements at 900 $^\circ$C in evacuated silica tubes. The crystals of bulk TaCoTe$_2$ are monoclinic with space group $P2_1/c$ (No.~14) and the crystal structure have been determined by the X-ray powder diffraction method (with crystal structure data $a = 7.7945$ \AA, $b = 6.2649$ \AA, $c = 8.1524$ \AA)~\cite{li1993x}. Our calculation shows that the bulk TaCoTe$_2$ is an AFM semimetal. In the absence of SOC, it hosts a nodal loop near the Fermi level, as shown in Appendix B.

Because of the layered structure, one expects that the monolayer TaCoTe$_2$ can be exfoliated from the bulk like other 2D materials such as graphene and MoS$_2$. The lattice of the monolayer owns the same space group symmetry as the bulk [see Fig.~\ref{fig1}(b)], which can be generated by the following elements: the inversion $\mathcal{P}$ and the glide mirror $\widetilde{\mathcal{M}}_{y}: (x,y,z)\rightarrow (x+\frac{1}{2},-y+\frac{1}{2},z)$. The combination of the
two operations leads to a screw axis $\widetilde{\mathcal{C}}_{y}: (x,y,z)\rightarrow (-x+\frac{1}{2},y+\frac{1}{2},-z)$. Here the tilde denotes a nonsymmorphic operation, which involves a fractional lattice translation. The fully relaxed monolayer structure has lattice parameters $a = 7.7746$ \AA\ and $b = 6.3374$ \AA. The details of our first-principles calculation are presented in the Appendix A.

To investigate the stability of the monolayer structure, we perform the phonon spectrum calculation. The obtained phonon spectrum is plotted in Fig.~\ref{fig3}(b) One observes that there is no soft mode in the spectrum, showing that the structure is dynamically stable.

After confirming the stability of monolayer TaCoTe$_2$, we calculate its exfoliation energy, which offers an indication of the easiness to obtain the monolayer from the bulk material by exfoliation process [see the inset of Fig.~\ref{fig3}(c)]. The exfoliation process is simulated by calculating the energy variation ($\delta E$) when a monolayer is separated from the bulk by a distance $d$. The energy saturates to a value with increasing $d$, which corresponds to the exfoliation energy. As shown in Fig.~\ref{fig3}(c), the exfoliation energy for monolayer TaCoTe$_2$ is about 0.406 J/m$^2$. This value is comparable to that of graphene (0.37 J/m$^2$)~\cite{zacharia2004} and MoS$_2$ (0.41 J/m$^2$), and is less than that of Ca$_2$N (1.14 J/m$^2$)~\cite{zhao2014,guan2015}. We have also calculated the exfoliation strength $\sigma$, which is defined as the maximum derivative of $\delta E$ with respect to the separation $d$. The obtained exfoliation strength is about 1.64 GPa, which is less than that of the graphene ($\sim$2.1 GPa)~\cite{zhao2014}. These results suggest that monolayer TaCoTe$_2$ should be readily obtained from its bulk material by mechanical exfoliation.

\section{Magnetic Configuration}

\begin{figure}[t]
\includegraphics[width=8cm]{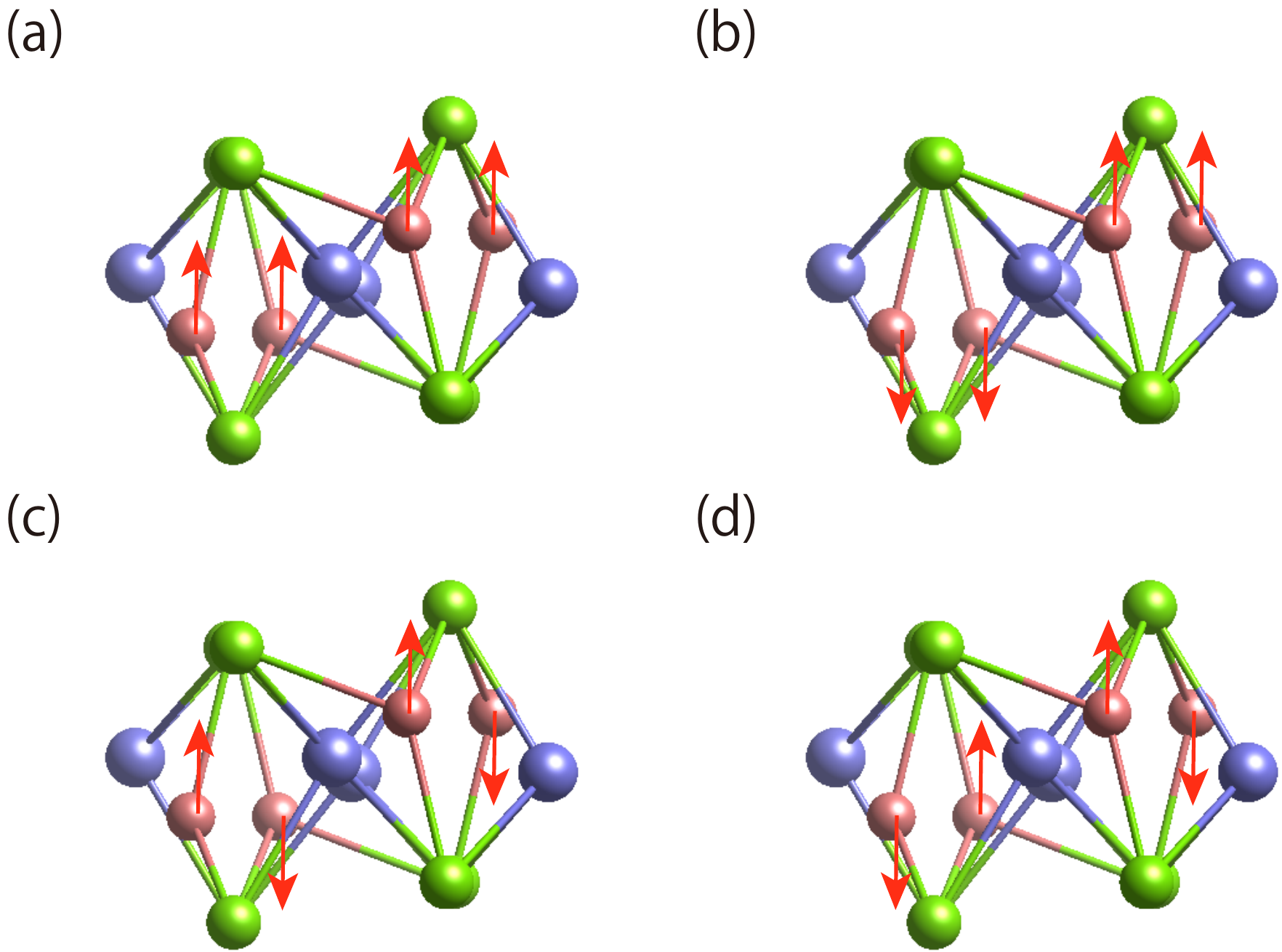}
\caption{The magnetic configurations that we have considered: (a) is for the ferromagnetic configuration, and (b)-(d) are for the three kinds of AFM configurations (AFM1-AFM3). The red arrows represent the direction of the magnetic moments}
\label{fig2}
\end{figure}

Because monolayer TaCoTe$_2$ contains the $3d$ transition metal element Co which usually exhibits magnetism, we shall first determine the magnetic ground state for the material. We consider the ferromagnetic (FM) configuration and three antiferromagnetic configurations (denoted as AFM1, AFM2, and AFM3), as illustrated in Figs.~\ref{fig2}(a-d). For each of these states, we consider three orientations for the magnetic moments, namely, the $z$, $x$, and $y$ directions ($z$ is the out-of-plane direction). Our first-principles calculations with GGA+SOC+$U$ approach show that the lowest energy is obtained for AFM1 with magnetic moments along the $z$ direction (denoted as AFM1$_z$). The comparison of the total energies for these magnetic configurations are shown in Table.~\ref{table1}. The relatively large energy difference between the FM and the AFM$1_z$ states indicates the stability of the AFM configuration.

\begin{table*}
	\caption{\label{table1} Total energy $E$ per unit cell (in eV, relative to that of the AFM$1_z$ ground state) and magnetic moment $M$ (in $\mu_B$) per Co atom obtained from first-principles calculations for the different magnetic configurations as illustrated in Fig.~\ref{fig2}.}
	\begin{ruledtabular}
		\begin{tabular}{lcr}
			\quad\qquad FM$_z$\quad AFM1$_z$\quad AFM2$_z$\quad AFM3$_z$\qquad\quad FM$_x$\quad AFM1$_x$\quad AFM2$_x$\quad AFM3$_x$\qquad\quad FM$_y$\quad AFM1$_y$\quad AFM2$_y$\quad AFM3$_y$\\
			\hline
			$E$ \ \quad 0.371\quad \ \ 0.000\qquad 0.323\qquad 0.030\qquad\quad  \ \  0.367\quad \  0.005\quad \ \ 0.335\qquad 0.035\qquad\qquad     0.352\quad \ 0.009\qquad 0.330\qquad 0.034\qquad\\
			$M$   \quad 0.886\quad \ \ 1.246\qquad 0.958\qquad 1.232\qquad\quad \ \  0.945\quad \ 1.238\quad \ \ 0.926\qquad 1.215\qquad\qquad     0.992\quad \ 1.241\qquad 0.914\qquad 1.234\qquad
		\end{tabular}
	\end{ruledtabular}
\end{table*}

\begin{figure}
	\includegraphics[width=8.4cm]{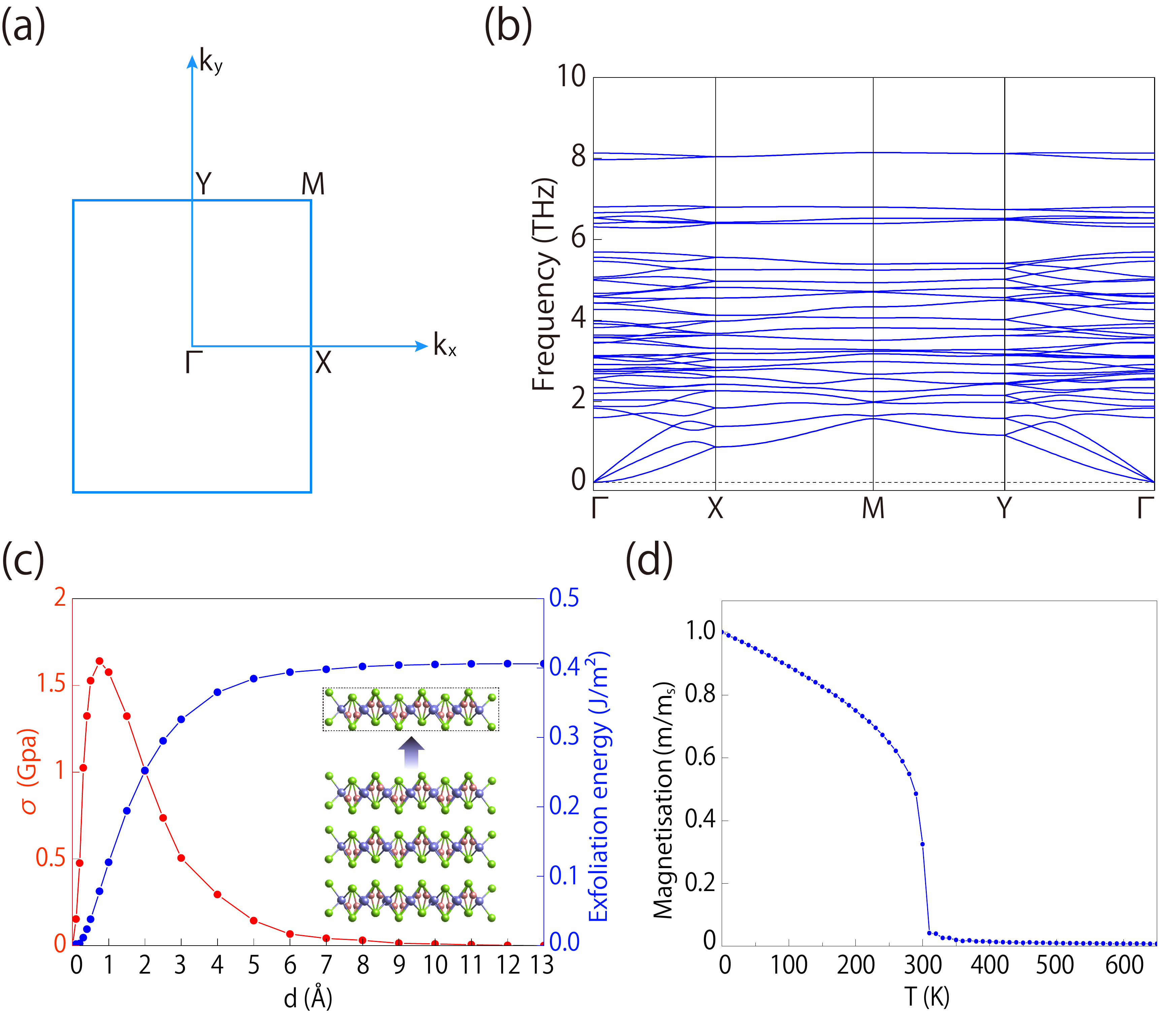}
	\caption{(a) Brillouin zone for the monolayer TaCoTe$_2$. The high-symmetry points are labeled. (b) Phonon spectrum of the monolayer TaCoTe$_2$. (c) Exfoliation energy (blue line) for TaCoTe$_2$ as a function of its separation distance $d$ from the bulk (as illustrated in the inset). Here the bulk is modeled by three TaCoTe$_2$ layers in the calculation. The red curve shows the exfoliation strength $\sigma$ (i.e., the derivative of exfoliation energy with respect to $d$). (d) Variation of the staggered magnetization $M_s$ with respect to the temperature. }
	\label{fig3}
\end{figure}

We have also estimated the N\'{e}el temperature ($T_N$) for the AFM$1_z$ ground state. The estimation is through the Monte Carlo (MC) simulation approach based on the effective spin model~\cite{evans2014atomistic}:
\begin{equation}\label{Heisenberg}
H=-\sum_{i \neq j} J_{i j} \bm{S}_{i} \cdot \bm{S}_{j}-\frac{k_{N}}{2} \sum_{i}\left(S_{i}^{z}\right)^{2},
\end{equation}
where $i$ and $j$ represent the Co atomic sites, $J_{ij}$ and $k_N$ are the exchange interaction strength and the magnetic anisotropy strength, respectively. The model parameters can be extracted from the first-principles calculations. The obtained nearest neighbor exchange interaction $J_1=-2.974\times 10^{-20}$ J and the next nearest neighbor $J_2=6.469\times 10^{-21}$ J. The magnetic anisotropy strength is $k_N=3.6\times 10^{-22}$ J.
The N\'{e}el temperature is determined from the variation of mean sublattice magnetization with respect to the temperature, which is shown in Fig.~\ref{fig3}(d) from the MC simulation. The estimated $T_N$ value is about 310 K. The relatively high N\'{e}el temperature suggests that the monolayer TaCoTe$_2$ can be suitable for practical spintronics applications.

\section{Without SOC: type-I AFM Dirac semimetal}

In the following, we investigate the electronic band structure for the ground state of monolayer TaCoTe$_2$ (i.e., with AFM$1_z$ configuration).

\begin{figure}[h]
	\includegraphics[width=7.8cm]{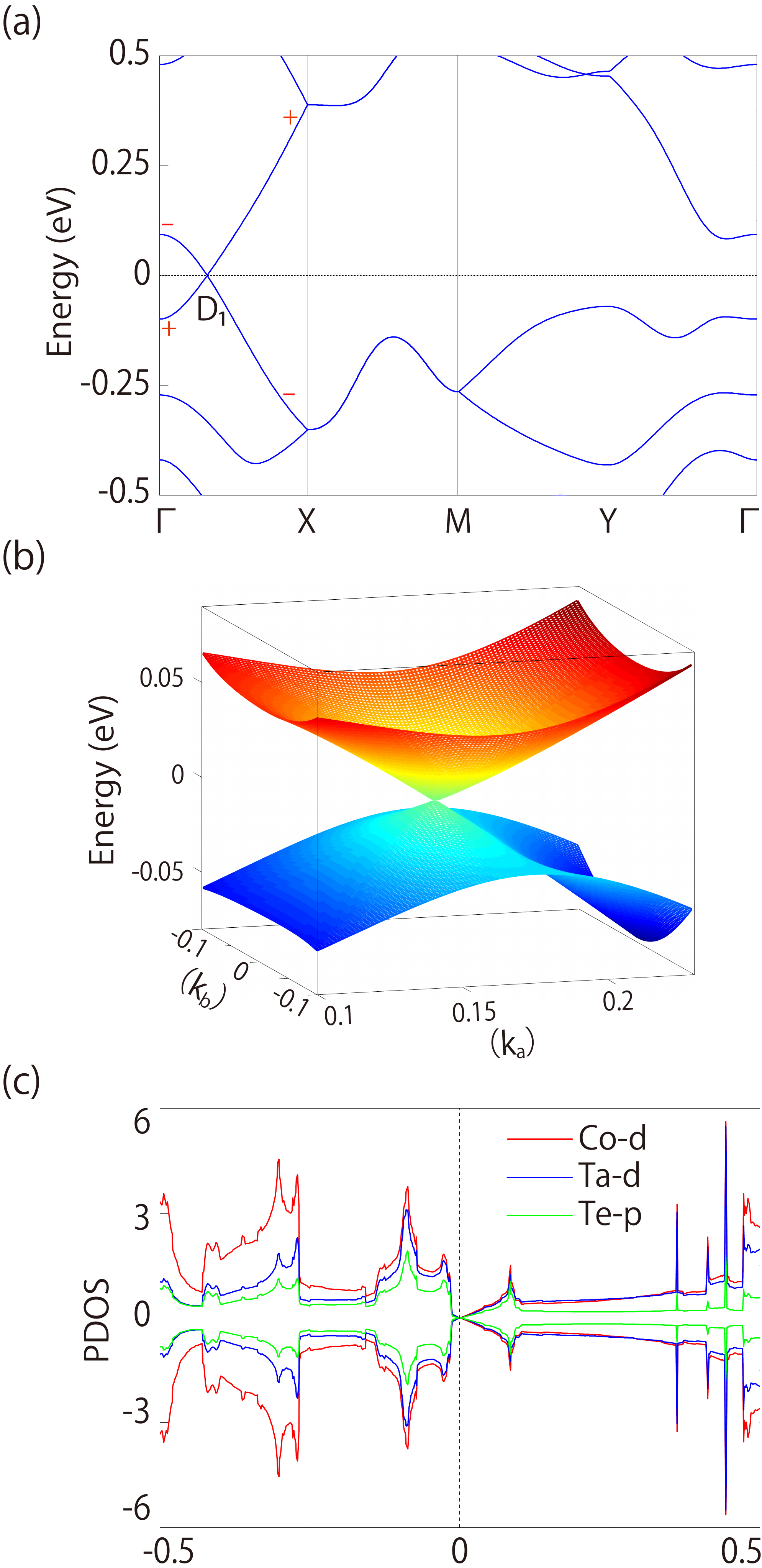}
	\caption{(a) Band structure of the monolayer TaCoTe$_2$ without SOC. The glide mirror $\widetilde{\mathcal{M}}_{y}$ eigenvalues for the two low-energy bands along $\Gamma$-X are labeled. (b) 2D band structure around the Dirac point $D_1$. (c) shows the projected density of states (PDOS).}
	\label{fig4}
\end{figure}

We first consider the band structure in the absence of SOC.
The band structure and the projected density of states (PDOS) from our calculation are shown in Fig.~\ref{fig4}. From the PDOS, one can see that the system is a semimetal: it has zero band gap and the density of states at the Fermi level vanishes in a linear manner as in graphene. The low-energy states near the Fermi level are mainly from the Co ($3d$), Ta ($5d$), and Te ($5p$) orbitals.
In the band structure, one observes that there is a linear band-crossing point $D_1$ along the $\Gamma$-X path exactly at the Fermi level (There is a pair of $D_1$ points symmetric about $\Gamma$.). In Fig.~\ref{fig4}(b), we plot the band dispersion around this crossing point, which demonstrates that the point is isolated and has linear dispersion.
Therefore, monolayer TaCoTe$_2$ in the absence of SOC is a 2D AFM Dirac semimetal, with massless Dirac fermion excitations.

The Dirac point here is protected by the glide mirror symmetry $\widetilde{\mathcal{M}}_{y}$. As shown in Fig.~\ref{fig4}(a), the two crossing bands have opposite $\widetilde{\mathcal{M}}_{y}$ eigenvalues along the $\Gamma$-X path. At $\Gamma$ and $X$ points, the ordering between the two bands are switched.
Thus, the two bands must linearly cross without hybridization at a Dirac point $D_1$ on the path. In Sec.VI, we shall further characterize the low-energy band structure around $D_1$ by constructing an effective model.

\section{With SOC: type-II AFM Dirac point}

\begin{figure}[h]
	\includegraphics[width=7.6cm]{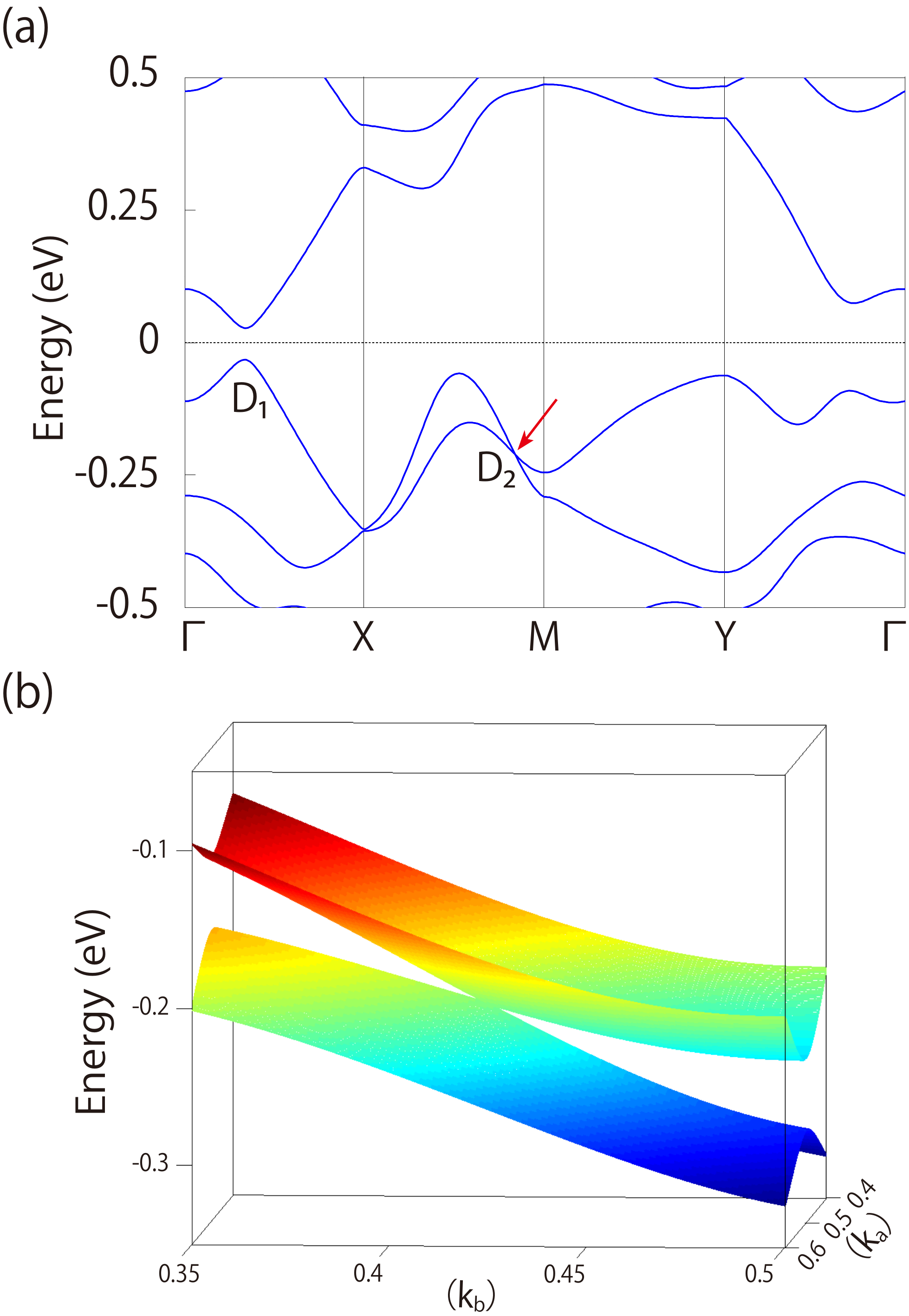}
	\caption{(a) Band structure of the monolayer TaCoTe$_2$ with SOC included. The type-II Dirac point $D_2$ is indicated by the red arrow. (b) 2D band structure around the type-II Dirac point $D_2$.}
	\label{fig5}
\end{figure}

Next, we turn to the band structure with SOC included. The DFT result is plotted in Fig.~\ref{fig5}(a). Here, each band is doubly degenerate because of the existence of the combined $\mathcal{PT}$ symmetry (although the $\mathcal{P}$ and $\mathcal{T}$ are individually broken by the magnetic order). One can observe that original Dirac point $D_1$ is gapped out (with a small band gap $\sim 60$ meV). This is because with SOC, the spatial symmetry operation also affects spin, and $\widetilde{\mathcal{M}}_{y}$ is not preserved for the AFM$1_z$ state, such that $D_1$ is no longer protected. As a result, the low-energy electrons become massive Dirac fermions.

More interestingly, one notes that there emerges a linear crossing point $D_2$ in the valence band about 0.2 eV below the Fermi level. In Fig.~\ref{fig5}(b), we show a zoom-in plot of the band structure around this crossing point, which confirms that this is an isolated Dirac point in 2D. Furthermore, this Dirac cone is tipped over, namely, the slopes of the two crossing bands have the same sign along the X-M ($k_y$) direction. This kind of Dirac point is termed as the type-II Dirac point, in analogy with the similar definition for the Weyl point. It has been shown that type-II Dirac/Weyl fermions can exhibit unique magnetic, optical, and transport properties distinct from their type-I counterparts. Previously, type-II Dirac points have been reported in a few nonmagnetic systems, such as PtTe$_2$~\cite{yan2017lorentz}, PtSe$_2$~\cite{huang2016type,zhang2017experimental}, PdTe$_2$~\cite{noh2017experimental,fei2017nontrivial} and $MA_{3}$ ($M=$ V, Nb, Ta; $A=$ Al, Ga, In)~\cite{chang2017type}. To our knowledge, this is the first time that a type-II Dirac point is revealed in a 2D magnetic material.

In the following, we clarify the symmetry protection for this Dirac point $D_2$. It is important to note that besides $\mathcal{PT}$, the magnetic configuration also preserves the $\widetilde{\mathcal{C}}_{y}$ symmetry. The path X-M at $k_x=\pi$ (below, wave vectors are measured in units of the respective inverse lattice parameter) is an invariant subspace for $\widetilde{\mathcal{C}}_{y}$, so each energy eigenstate on this path can also be chosen as an eigenstate of $\widetilde{\mathcal{C}}_{y}$. On X-M, we have
\begin{equation}
\widetilde{\mathcal C}_y^2=T_{01}\overline{E}=-e^{-ik_y},
\end{equation}
where $T_{01}$ denotes a translation along $y$ by a lattice constant, and $\overline{E}$ denotes the $2\pi$ spin rotation. Hence, the eigenvalues of $\widetilde{\mathcal C}_y$ are given by $s=\pm i e^{-ik_y/2}$.
Meanwhile, the commutation relation between $\widetilde{\mathcal{C}}_{y}$ and $\mathcal{PT}$ on X-M is given by
\begin{equation}\label{CyPT}
\widetilde{\mathcal{C}}_{y}\mathcal{PT}=T_{11}\mathcal{PT}\widetilde{\mathcal{C}}_{y}=-e^{-ik_{y}}\mathcal{PT}\widetilde{\mathcal{C}}_{y}.
\end{equation}
Using Eq.~(\ref{CyPT}), one finds that for any state $|u \rangle$ with eigenvalue $s$, its Kramers partner $\mathcal{PT}|u\rangle$ satisfies
\begin{equation}\label{eq3}
\widetilde{\mathcal{C}}_{y}(\mathcal{P}\mathcal{T}| u \rangle)
=s(\mathcal{P}\mathcal{T}| u \rangle).
\end{equation}
This shows that the degenerate pair $|u \rangle$ and $\mathcal{P}\mathcal{T}| u \rangle$ have the same $\widetilde{\mathcal{C}}_{y}$ eigenvalue $s$. It follows that if two bands (each is $\mathcal{P}\mathcal{T}$ doubly degenerate) have opposite $s$ on X-M, their crossing point would be a protected Dirac point with fourfold degeneracy. This is exactly the case for the point $D_2$ here. Deviating from the X-M path, the $\widetilde{\mathcal{C}}_{y}$ protection is lost, so $D_2$ must be an isolated Dirac point.


\section{effective model}

In the following, we construct effective models to describe the electronic states around $D_1$ and $D_2$. Following Tang \emph{et al.}~\cite{Tang2016}, the effective four-band model around a band crossing point for a $\mathcal{PT}$ symmetric system takes the general form of
\begin{equation}\label{kp1}
\begin{split}
{\mathcal{H}}(\boldsymbol{k})=&d_0(\boldsymbol{k})+d_1(\boldsymbol{k})\tau_x+d_2(\boldsymbol{k})\tau_z+d_3(\boldsymbol{k})\tau_y\sigma_x\\
&+d_4(\boldsymbol{k})\tau_y\sigma_y+d_5(\boldsymbol{k})\tau_y\sigma_z,
\end{split}
\end{equation}
where $\tau$'s ($\sigma$'s) are Pauli matrices representing the orbital (spin-related AFM) basis, and $d_i(\bm k)$ ($i=0,1,\cdots,5$) are real functions of $\bm k$. By writing down this form, we have taken the representation $\mathcal{PT}=i\sigma_y K$, where $K$ is the complex conjugation operator.

Let's first consider the model for $D_1$. Without SOC, the $d_3$ and $d_4$ terms in (\ref{kp1}) must vanish, because they correspond to spin-flip processes. The bands cross at $D_1$ on the $\Gamma$-X path and they have opposite $\widetilde{\mathcal{M}}_{y}$ eigenvalues in the absence of SOC. It follows that $\widetilde{\mathcal{M}}_{y}=e^{-ik_x/2}\tau_z$, and the Dirac model constrained by symmetry around $D_1$ can be obtained to linear order as
\begin{equation}
  \mathcal{H}_{D1}^0(\bm k)=v_0 k_x +v_1 k_x \tau_z +v_2 k_y \tau_x +v_3 k_y\tau_y\sigma_z,
\end{equation}
where the wave vector $\boldsymbol{k}$ is measured from $D_1$, and the $v$'s are real model parameters. Consequently, the band dispersion around $D1$ is given by
\begin{equation}
  E(\bm k)=v_0 k_x\pm\sqrt{v_1^2 k_x^2 +(v_2^2+v_3^2)k_y^2},
\end{equation}
which describes the massless Dirac cone in the absence of SOC.

When SOC is turned on, there are additional allowed terms. By treating SOC as a perturbation and keeping only the leading order terms, we have
\begin{equation}
   \mathcal{H}_{D1}(\bm k)=\mathcal{H}_{D1}^0+\lambda_1\tau_y\sigma_x+\lambda_2 \tau_y\sigma_y +\lambda_3 \tau_y\sigma_z,
\end{equation}
where the $\lambda$'s are real parameters.
Clearly, the additional SOC terms open up a gap $\Delta=2\sqrt{\lambda_1^2+\lambda_2^2+\lambda_3^2}$ at $D_1$,  and make the Dirac fermions massive.

Next, we consider the model for the $D_2$ point. As we have discussed before, on the path X-M, the symmetry $\widetilde{\mathcal{C}}_{y}$ is preserved. The exact representation of the $\widetilde{\mathcal{C}}_{y}$ is obtained by noticing
\begin{equation}
\widetilde{\mathcal{C}}_{y}\mathcal{PT}=T_{11}\mathcal{PT}\widetilde{\mathcal{C}}_{y}=e^{-ik_{x}} e^{-ik_{y}}\mathcal{PT}\widetilde{\mathcal{C}}_{y}.
\end{equation}
Thus $\widetilde{\mathcal{C}}_{y}=i e^{-ik_y/2}\tau_z$ when $k_x=\pi$. Consequently, in the presence of SOC, the effective model for $D_2$ is given by
\begin{equation}\label{kp4}
\begin{split}
\mathcal{H}_{D2}(\boldsymbol{\mathit{q}})=&w_{0}q_{y}+w_{1}q_{x}\tau_{x}+w_{2}q_{y}\tau_{z}+w_{3}q_{x}\tau_{y}\sigma_{x}\\
&+w_{4}q_{x}\tau_{y}\sigma_{y}+w_{5}q_{x}\tau_{y}\sigma_{z},
\end{split}
\end{equation}
where the energy and the wave vector $\bm q$ is measured from $D_2$.
The first term $w_{0}q_{y}$ represents the tilt term. Since $D_2$ is a type-II Dirac point, we have $\left|w_{0}\right|>\left|w_{2}\right|$, such that the tilt dominates the dispersion along $q_y$ and makes the Dirac cone tipped over.

\section{control Dirac point via magnetism}

\begin{figure*}
	\includegraphics[width=15cm]{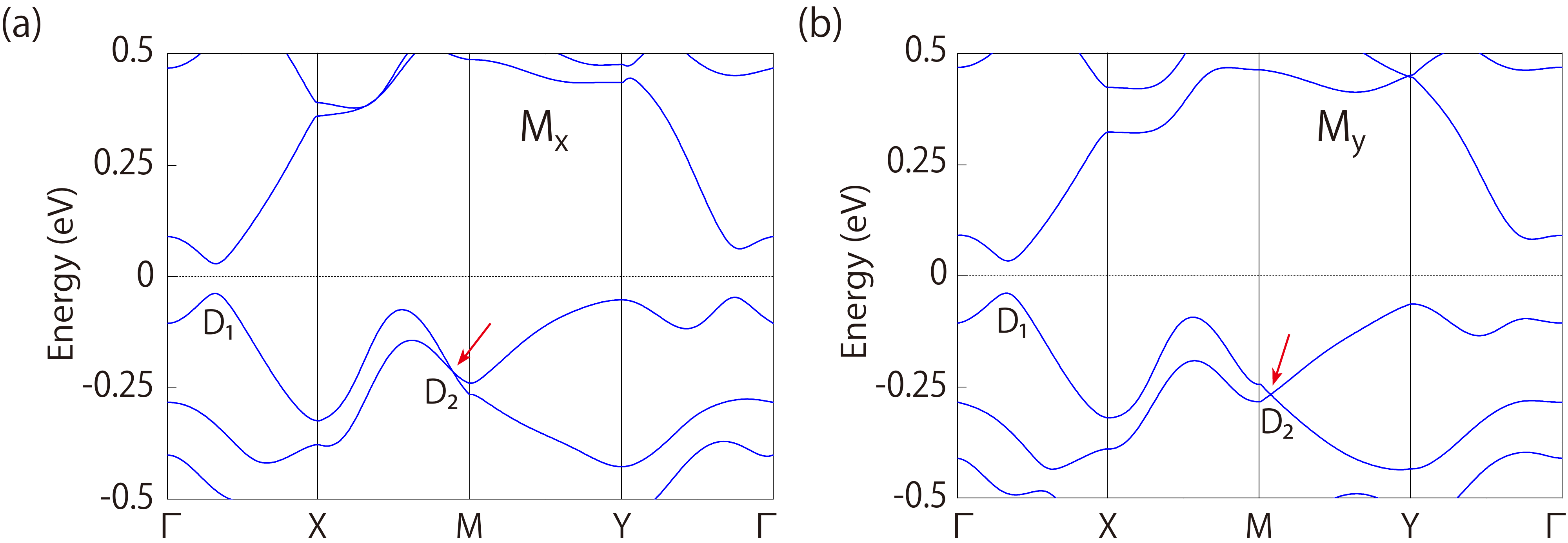}
	\caption{Band structures of the monolayer TaCoTe$_2$ (in the present of SOC) when the N\'{e}el vector is in the (a) $x$ and (b) $y$ direction. The red arrows indicate the AFM Dirac point. It is of type-II in (a) and of type-I in (b).}
	\label{fig6}
\end{figure*}

We have discussed the type-II Dirac point which occurs in the AFM1$_z$ ground state, where the N\'{e}el vector $\bm n$ is along the $z$ direction. When SOC is considered, the electronic band structure will be affected by the orientation of the N\'{e}el vector.

In Fig.~\ref{fig6}, we show the results when $\bm n$ is along $x$ and $y$ directions. One can see that when $\bm n$ is along the $x$ direction, there still exists a Dirac point on the X-M path, although the band structure slightly changes from Fig.~\ref{fig5}(a). This can be understood by noticing that this configuration shares the same symmetry with AFM1$_z$. Interestingly, for $\bm n$ along the $y$ direction, although the Dirac point on X-M disappears,
there emerges a Dirac point on the M-Y path (see Fig.~\ref{fig6}(b)), and it is a type-I Dirac point. This Dirac point is protected by $\mathcal{PT}$ and the
$\widetilde{\mathcal{M}}_{y}$ symmetry on this path. The analysis is very similar to that in Sec.V, so we will not repeat it here.

For $\bm n$ along other directions, one generally does not expect a stable Dirac point, because a stable Dirac point in 2D must require certain crystalline symmetry protection.

Experimentally, the N\'{e}el vector for a AFM can be rotated by applied magnetic field, laser light, and exchange bias~\cite{baltz2018}. Therefore, the result here indicates that magnetic Dirac points in 2D can be effectively tuned (including both location and dispersion) by controlling the magnetism.

\section{Discussion and Conclusion}

We have revealed monolayer TaCoTe$_2$ as a fertile playground to study magnetic Dirac points in 2D. The low-energy electrons are 2D Dirac fermions. In the absence of SOC, they are massless. The SOC effect endows them with a finite mass. 2D Massive Dirac fermions can exhibit many interesting physics. For example, they can acquire finite Berry curvature and orbital magnetic moment, leading to exotic transport and optical properties. In addition, similar to graphene and 2D transition metal dichalcogenides, there are two valleys in the band structure, which may lead to interesting valleytronic applications.

Notably, the type-II magnetic Dirac point found here has not been reported before. The type-II dispersion has Fermi surface topology distinct from type-I, which in turn results in many distinct physical effects. For example, there can be zero-field magnetic breakdown and peculiar magneto-optical properties for type-II dispersion~\cite{Yu2016a}. The tipped-over Dirac cone mimics the case inside a black hole in general relativity, hence it may give rise to effects analogous to event horizons and Hawking radiation~\cite{guan2017artificial,huang2018black}.

Finally, we comment on the experimental aspect. As we have discussed, bulk TaCoTe$_2$ has already been synthesized, and the monolayer should be readily obtained via mechanical exfoliation method. The key Dirac features in the band structure can be directly imaged by current ARPES technique. The type-II Dirac point is in the valence band and close to Fermi level, which is very suitable for ARPES detection. In the intrinsic case, the transport property of monolayer TaCoTe$_2$ should be dominated by the massive Dirac fermions at the band edges. To probe the type-II Dirac point in transport experiment, one needs to tune the Fermi level down into the valence band, which can be achieved by electric or ionic gating.

In conclusion, we have revealed the monolayer TaCoTe$_2$ as a 2D AFM Dirac material. Our first-principles calculation shows that the material is stable and can be readily obtained from its bulk counterpart. The ground state is an AFM, and it has a relatively high N\'{e}el temperature.
In the absence of SOC, the materials is an ideal 2D AFM Dirac semimetal, with a pair of magnetic Dirac points at the Fermi level. With SOC, the low-energy Dirac fermions become massive, and meanwhile, there emerges a pair of type-II magnetic Dirac points in the valence band close to the Fermi level. We show that the Dirac points can be tuned by controlling the orientation of the N\'{e}el vector. Our results here offer an excellent platform for exploring the intriguing physics of 2D magnetic Dirac fermions, which can lead to potential applications in AFM spintronics.

\begin{acknowledgements}
The authors thank D. L. Deng for valuable discussions. {The work is supported by the NSF of China (Grants No.~11734003 and 11574029), the National Key R$\&$D Program of China (Grant No. 2016YFA0300600), the Strategic Priority Research Program of Chinese Academy of Sciences (Grant No. XDB30000000).}, and the Singapore Ministry of Education AcRF Tier 2 (Grant No.~MOE2017-T2-2-108).
\end{acknowledgements}

\begin{appendix}

\section{First-principles Methods}

Our first-principles calculations are based on the density functional theory (DFT) using the projector augmented wave method as implemented in the Vienna ab initio simulation package~\cite{Kresse1994,Kresse1996,PAW}. The exchange-correlation functional was modeled within the generalized gradient approximation (GGA) with the Perdew-Burke-Ernzerhof (PBE) realization~\cite{PBE}. The cutoff energy was set as 400 eV, and a $9\times 9\times 1$ $\Gamma$-centered $k$-point mesh was used for the Brillouin zone sampling. The energy and force convergence criteria were set to be $10^{-5}$ eV and $0.01$ eV/\AA, respectively. A vacuum layer with a thickness of 20 \AA\, was taken to avoid artificial interactions between periodic images. The phonon spectrum is calculated using the PHONOPY code through the DFPT approach~\cite{Togo2015}. To account for the correlation effects for the electrons on Co-3$d$ orbitals, the DFT$+U$ method~\cite{Anisimov1991,dudarev1998} was used for calculating the band structures. For the results presented in the main text, the $U$ value was taken to be 3 eV. The test of other $U$ values is presented in Appendix C. 

\section{Band structure of the bulk TaCoTe$_2$}
\begin{figure*}
	\includegraphics[width=14cm]{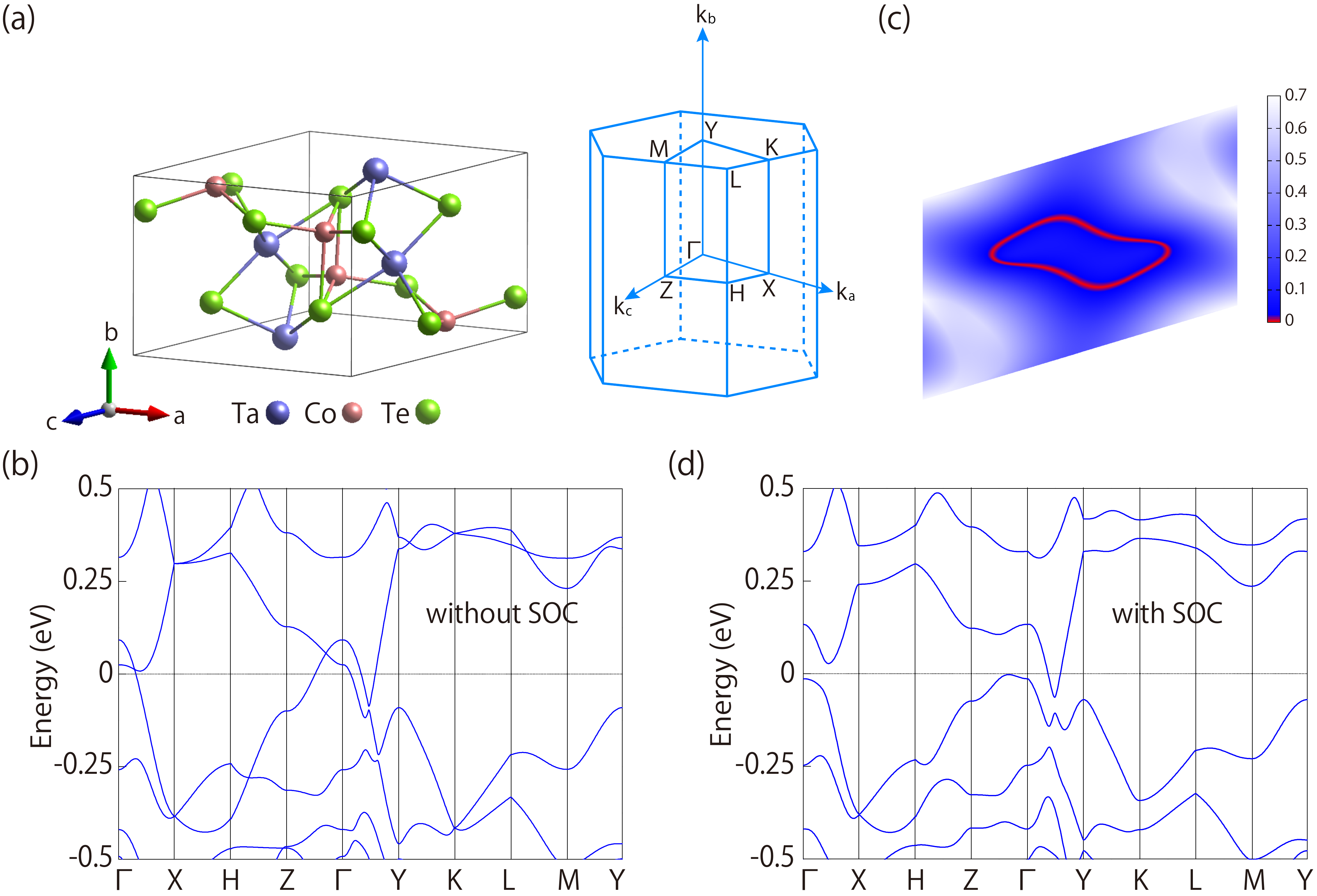}
	\caption{(a) Unit cell and Brillouin zone for bulk TaCoTe$_2$. (b) Band structure of bulk TaCoTe$_2$ without SOC. (c) Shape of the nodal loop (red curve) obtained from the DFT calculation. The color map indicates the local gap between the two crossing bands. (d) Band structure of bulk TaCoTe$_2$ with SOC included.}
	\label{fig7}
\end{figure*}

Our first-principles calculations show that the ground state of the bulk TaCoTe$_2$ share the same AFM configuration (AFM$1_z$) within each monolayer (For the calculation of the bulk, we adopted the experimental lattice parameters and a $9\times 9\times 8$ $\Gamma$-centered $k$-point mesh. The van der Waals (vdW) corrections have been taken into
account by the approach of Dion et al.~\cite{Dion2004}). Figure~\ref{fig7} shows the band structures of the bulk without and with SOC. In the absence of SOC, there exist linear band-crossing
points along the $\Gamma$-X and Z-$\Gamma$ paths near the Fermi
level [see Fig.~\ref{fig7}(b)]. We find
that the two points are not isolated and they belong to a nodal loop centered around the $\Gamma$ point in the $k_b$ = 0 plane, as shown in Fig.~\ref{fig7}(c). When SOC is considered, a gap is opened at the nodal loop [see Fig.~\ref{fig7}(d)].

\section{Band structure results with different Hubbard U correction values}
\begin{figure*}
\includegraphics[width=17cm]{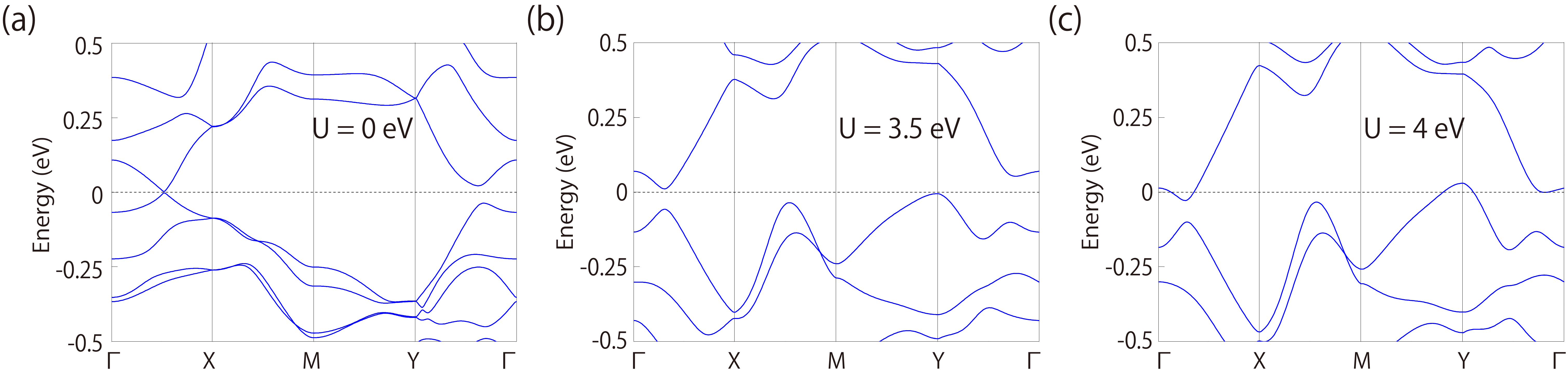}
\caption{Band structure results of the monolayer TaCoTe$_2$ with different $U$ values: (a) 0 eV, (b) 3.5 eV, and (c) 4 eV. The SOC is included.}
\label{fig8}
\end{figure*}

In order to observe how the band structure of the monolayer TaCoTe$_2$ vary with the Hubbard $U$ parameter. We have tested different $U$ values with SOC included. The representative results are displayed in Fig.~\ref{fig8}. We can see that with small $U$ correction ($U=0$ eV), the valence band and conduction band nearly touch and there is almost no gap at $D_1$ (with SOC included). With large $U$ values (such as 4 eV), the global gap is reduced and finally closed, but a local gap still exist. The type-II Dirac point $D_2$ in the valence band remains robust.

\end{appendix}



%

\end{document}